# Influences of Electroosmotic Flow on Ionic Current through Nanopores: a Comprehensive Understanding

Yinghua Qiu (裘英华),[1, 2, 3, 4]* and Long Ma (马龙)[1]

1. Key Laboratory of High Efficiency and Clean Mechanical Manufacture of Ministry of Education, National Demonstration Center for Experimental Mechanical Engineering Education, School of Mechanical Engineering, Shandong University, Jinan, 250061, China

2. Shenzhen Research Institute of Shandong University, Shenzhen, Guangdong, 518000, China

3. Suzhou Research Institute of Shandong University, Suzhou, Jiangsu, 215123, China

4. Key Laboratory of Ocean Energy Utilization and Energy Conservation of Ministry of Education, Dalian, Liaoning, 116024, China

*Corresponding author: yinghua.qiu@sdu.edu.cn

**ABSTRACT**

Continuum simulations become an important tool to uncover the mysteries in nanofluidic experiments. However, fluid flow in simulation models is usually unconsidered. Here, systematical simulations are conducted to provide a quantitative understanding of influences from electroosmotic flow (EOF) on ionic transport through nanopores by both types of models with and without consideration of EOF. In nanopores of less than ~10 nm in diameter, counterions dominate ionic current which is always promoted obviously by the convective effect of EOF. In the diameter range from ~10 to ~30 nm, strong EOF induces ion concentration polarization or ion depletion inside nanopores which causes significant decreases in ionic current. For nanopores larger than ~30 nm, due to convective promotion and inhibition of EOF on the transport of counterions and anions, considerable nanopore selectivity to counterions maintains in cases with EOF. Though the difference in total current between both cases decreases with further pore size increasing, the difference in cation/anion current is still considerable. From our results under various pore parameters and applied conditions, the fluid flow should be considered in the simulation cases when EOF is strong. Our work may provide useful guidance for simulation conductance.

**I. INTRODUCTION**

Nanopores provide a versatile platform to investigate the transport of ions and water molecules.[1-7] Under such highly confined spaces, the mass transport through nanopores can be tuned effectively by the properties of pore walls and solution conditions. Many important electrokinetic phenomena have been discovered with various nanopores, such as electro-osmosis/electrophoresis,[1] stream potential/current,[8] ionic current rectification,[3] ion concertation polarization,[9,10] resistive pulses,[11] and surface-charge-governed ionic conductance,[12,13] which may find numerous applications in biosensing,[14] ionic circuits and amplifiers,[5] desalination,[15] preconcentration of analytes,[16] electroosmotic pumps,[17] and energy conversion.[18]

In the above cases, due to electrostatic interactions between surface charges and free ions, electric double layers (EDLs) form spontaneously near charged pore walls.[19-21] Based on the hydration effect,[22] electroosmotic flow (EOF) is induced by the directional movement of counterions inside EDLs under applied electric fields. The EOF velocity can be influenced by the properties of nanopores and experimental conditions, such as the geometry[23,24] and surface charge of nanopores,[25,26] applied voltage,[27] as well as salt types[28,29] and concentration.[24]

Because of the symbiotic relationship between EOF and ionic movement, the investigation of the microscopic modulation mechanism of fluid flow on ionic transport is usually conducted with simulations.[2,3,30] Since EOF shares the same direction with counterions, EOF provides a convective contribution and inhibition to the movement of counterions and coions, respectively, which may significantly affect the dynamic and static characteristics of ions inside and outside of the nanopores. Through building both simulation models with and without consideration of fluid flow, Ai et al.[31] investigated the effect of EOF on ionic current rectification in conical nanopores. They found that obvious

EOF under high voltages or large surface charge densities can induce significant influences on ionic transport through nanopores. By simulating ionic flow in various electrolyte solutions i.e. different diffusion coefficients of cations, Hsu et al.[32] confirmed that the influences of EOF on the ionic current rectification could not be neglected through conical nanopores.

Simulation by the finite element method such as COMSOL Multiphysics provides an essential way to investigate the detailed ionic behaviors in nanopores. In continuum simulations, fluid flow through nanopores is described with the Navier-Stokes (NS) equations. However, due to various reasons such as the heavy workload of solving NS equations, or the weak fluid flow in highly confined spaces, EOF is unconsidered in many simulations.[33-41] Although the trend of experimental results was captured by simulations without EOF, large deviations from the real cases may appear when strong EOF exists such as in the cases with high applied voltages, low salt concentration, and highly charged nanopores. Because most electrokinetic applications depend on ionic transport, accurate simulation becomes especially important.[42]

Here, systematical simulations were conducted to provide a quantitative understanding of the EOF influence on the ionic transport through short cylindrical nanopores, due to the wide applications of short cylindrical nanopores in nanofluidic sensing of single molecules and nanoparticles. Because of the dependence of EOF speed on the pore dimension and applied conditions, the pore size and length, surface charge density, salt concentration, and voltage were considered to explore the influences of EOF on the ionic current through nanopores. In the cases with consideration of EOF, coupled Poisson-Nernst-Planck (PNP) and NS equations were solved. While only PNP equations were used in the simulation models without consideration of EOF. With the variation of the pore size, three different modulation mechanisms of EOF on ionic current have been found. For highly confined nanopores with sub-10 nm diameters, due to accumulated counterions, EOF provides obvious convective promotion to the transport of counterions. In nanopores with a diameter between ~10 to ~30 nm, EOF has the fastest speed which affects the ionic concentration obviously by ion concentration polarization or ionic depletion. The ionic current is significantly decreased by EOF. In the cases without EOF, nanopores wider than ~30 nm almost lose ionic selectivity. However, the appearance of EOF can maintain nanopores selective to counterions because EOF provides the convective promotion and inhibition to the transport of counterions and coions, respectively. As the pore size increases further, though the influence of EOF on the total current diminishes, its effect on cation/anion current is still considerable. Our results uncovered microscopic influences of the EOF on the ionic current and concentration distribution inside nanopores, which can provide a useful guide to the conduction of simulations.

## II. METHODOLOGY

Ionic transport through nanopores was investigated with nanofluidic simulations by COMSOL Multiphysics. As shown in Figure S1, two-dimensional axisymmetric simulation models were selected, in which the nanopore is sandwiched between two reservoirs of 5 μm in diameter and length. With equations 1-4,[43, 44] coupled Poisson-Nernst-Planck (PNP) and Navier-Stokes (NS) equations were solved to consider the ion distribution near charged pore walls, as well as the ionic transport and fluid movement through nanopores. For simulations without consideration of EOF, only PNP equations were solved.[2, 3]

$$\varepsilon \nabla^2 \varphi = -\sum_{i=1}^{N} z_i F C_i \tag{1}$$

$$\nabla \cdot \mathbf{J}_i = \nabla \cdot \left( C_i \mathbf{u} - D_i \nabla C_i - \frac{Fz_i C_i D_i}{RT} \nabla \varphi \right) = 0 \tag{2}$$

$$\mu \nabla^2 \mathbf{u} - \nabla p - \sum_{i=1}^{N} (z_i F C_i) \nabla \varphi = 0 \tag{3}$$

$$\nabla \cdot \mathbf{u} = 0 \tag{4}$$

where $\mathbf{J}_i$, $C_i$, $D_i$, and $z_i$ are their corresponding ionic flux, concentration, diffusion coefficient, and valence of ionic species $i$ ($K^+$ ions and $Cl^-$ ions), respectively. $p$, $F$, $R$, $T$, $N$, $\varphi$, $\mathbf{u}$, and $\varepsilon$ are the pressure, Faraday's constant, gas constant, temperature, number of ion species, electrical potential, velocity, and dielectric constant of solutions respectively.

From the literature, there are no well-developed theoretical predictions for the ionic current through nanopores. With equations 5-8, Ionic current ($I$) through nanopores is usually evaluated by the applied voltage ($V$) over the total resistance ($R$). There are two components in the total resistance, i.e. the pore resistance ($R_p$) and access resistance ($R_{ac}$).[4]

$$R = R_p + 2R_{ac} \tag{5}$$

$$R_p = \frac{4L}{\pi \kappa d^2} \tag{6}$$

$$R_{ac} = \frac{1}{2\kappa d} \tag{7}$$

$$I = V/R = \frac{V}{R_p + 2R_{ac}} \tag{8}$$

in which $\kappa$ is the conductivity of the solution. $d$ and $L$ are the diameter and length of the pore.

$$V_p = \frac{R_p}{R} V \tag{9}$$

$$E_p = \frac{V_p}{L} \tag{10}$$

From equations 9 and 10, the voltage ($V_p$) and electric field strength ($E_p$) of the nanopore can be obtained.

According to the Smoluchowski equation (equation 11),[45, 46] the velocity of EOF is strongly dependent on the zeta potential ($\zeta$), electrical field strength ($E_p$), permeability ($\varepsilon$), and viscosity ($\eta$) of the fluid.

$$v_{EOF} = -\frac{\varepsilon \zeta E_P}{\eta} \tag{11}$$

The zeta potential ($\zeta$) can be assumed to be approximately equal to the surface potential ($\varphi_0$).[46] Based on the Grahame equation (equation 12),[19] the relationship is obtained between the velocity of EOF and pore length, diameter, electric potential, surface charge density, and the concentration of solutions (equation 13).

$$\sigma = 0.117 \sinh(\varphi_0/51.4)\sqrt{C_b} \tag{12}$$

$$v_{EOF} \approx 205.6 \frac{\varepsilon V}{\eta(4L+\pi d)} \text{arcsinh}(\frac{\sigma}{0.117\sqrt{C_b}}) \tag{13}$$

where $\sigma$, $\varphi_0$, and $C_b$ are the surface charge density, surface potential, and KCl solution concentration, respectively.

To provide a comprehensive investigation of the influences of EOF on ionic current, various nanopore parameters and solution conditions had been considered. Nanopores used here varied from 2 to 1000 nm in diameter and 30 to 500 nm in length. The default length was selected as 100 nm. The surface charge density changed from −0.04 to −0.12 C/m$^2$ with its default value of −0.08 C/m$^2$. KCl was selected as the electrolyte with the bulk concentration varying from 15 to 300 mM of which 100 mM was the default concentration. Diffusion coefficients for K$^+$ and Cl$^-$ ions were 1.96×10$^{-9}$ and 2.03×10$^{-9}$ m$^2$/s, respectively. Applied voltages were used from 0.5 to 1.5 V across the nanopore. In the simulations, the temperature and dielectric constant of aqueous solutions were set to 298 K and 80. Table S1 and Figure S2 provide the detailed boundary conditions and meshing strategy, as used in our previous works.[2, 3, 30]

**III. RESULT AND DISCUSSIONS**

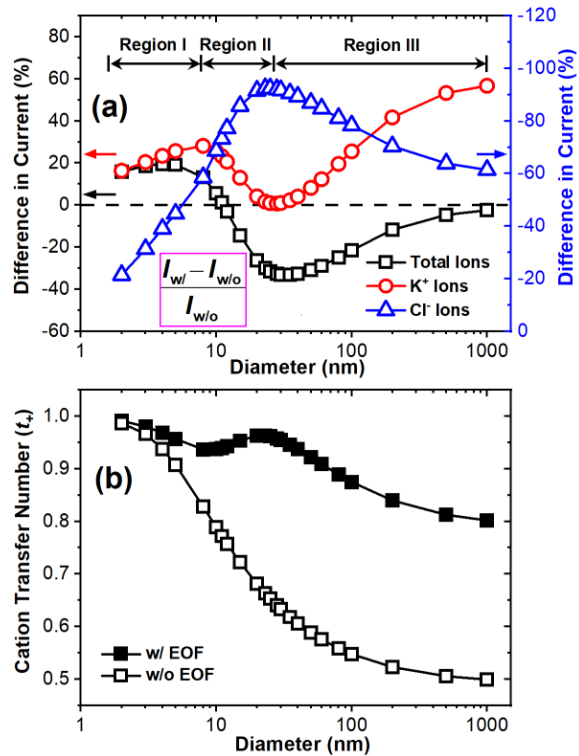

Figure 1 Effect of EOF on ionic behaviors through 100-nm-long nanopores with different diameters. Plotted data were the normalized difference in ionic current from cases with and without consideration of EOF. (a) Differences in ionic current contributed by K$^+$ ions, Cl$^-$ ions, and total ions. Three regions are defined as I, II, and III based on the profile of K$^+$ ions. (b) Transfer numbers of cation ($t_+$) from both cases. The surface charge density was −0.08 C/m$^2$. The solution and voltage were 100 mM KCl, and 1.0 V, respectively.

In aqueous solutions, ions are hydrated.[22] EOF results from the directional movement of counterions in EDLs under electric fields.[47] The thickness of EDLs near charged pore walls can be roughly evaluated by the Debye length which depends on the salt concentration.[19] Through varying the pore diameter (*d*) from 2 to 1000 nm, different confinement has been considered which tunes the overlapping degree of EDLs inside nanopores, as well as the length/diameter ratio. Because of the high throughput and sensitivity, thin nanopores find wide applications in nanofluidic sensing.[48-50] 100 nm is chosen as the default pore length.

In our simulation models, the fluid flow is described by NS equations. For the cases without consideration of EOF, only PNP equations were solved. With the ionic current obtained from both types of models, the difference in ionic current is analyzed with $\left(I_{w/} - I_{w/o}\right)/I_{w/o}$, in which $I_{w/}$ and $I_{w/o}$ represent the ionic current from cases with and without consideration of EOF. As shown in Figure 1a, the difference shows the influences of EOF on the ionic current through nanopores, which can be promotion or suppression depending on the pore size. When the diameter is less than ~10 nm, EOF has a promotion to the total ionic current,[51] which exhibits an increase-decrease profile and reaches the maximum value of ~20% at *d* ~4 nm. As the diameter exceeds 10 nm, EOF starts to suppress the ionic current, which also shows an increase-decrease trend with a maximum suppression of ~33% at *d* ~30 nm. With the diameter increasing further, this suppression effect of EOF on the ionic current gradually disappears.

Taking advantage of the simulation method, detailed ionic behaviors were investigated by analyzing the difference in ionic current contributed by anions or cations from cases with and without consideration of EOF. From Figure 1a, EOF always promotes and suppresses the current of $K^+$ ions and $Cl^-$ ions, respectively, which degree depends on the pore size.[32, 52] For $Cl^-$ ions, with the increase of the diameter the suppression first increases and then decreases, which reaches the maximum of ~95% at $d$ ~25 nm. The influences of EOF on the transport of cations are more complicated. Because of the large transfer number of cations in negatively charged nanopores (Figure 1b), we focus on the variation of $K^+$ ion current with the pore size, which can be divided into 3 regions. Region I, with a diameter from 2 to ~8 nm, there is enhanced promotion from ~16% to ~28% in the current of $K^+$ ions by EOF. Region II, for nanopores with a diameter between ~8 and ~25 nm, EOF provides a weakened promotion to $K^+$ ion current. The enhancement by EOF to the $K^+$ ion current gradually decreases from ~28% to ~0.9%. Region III, with the pore size varying from ~25 to ~1000 nm, the promotion of EOF to cation movement is enhanced to ~55%. The influences of EOF on cation transport can be inhibition in some cases discussed later.

The cation transfer number ($t_+$) shown in Figure 1b represents the ionic selectivity of nanopores to counterions.[2, 18] Due to the negatively charged pore walls, small nanopores exhibit perfect selectivity of counterions under both simulation conditions. In the cases without consideration of EOF, the cation transfer number decreases dramatically with the pore size. $t_+$ reduces from ~0.98 to ~0.7 as $d$ enlarges from ~2 to ~20 nm. With the diameter increasing further to ~100 nm, the nanopore almost loses the cation selectivity. However, in the cases with consideration of EOF, due to the significant promotion and inhibition of EOF in cation and anions transport, the cation transfer number is improved significantly.[31] For nanopores with diameters ranging from 2 to 70 nm, $t_+$ shows an oscillatory profile whose value keeps larger than 0.9. Also, with pore size increasing further, it has a much slower drop which remains a surprising value of ~0.8 even at $d$ ~1000 nm.

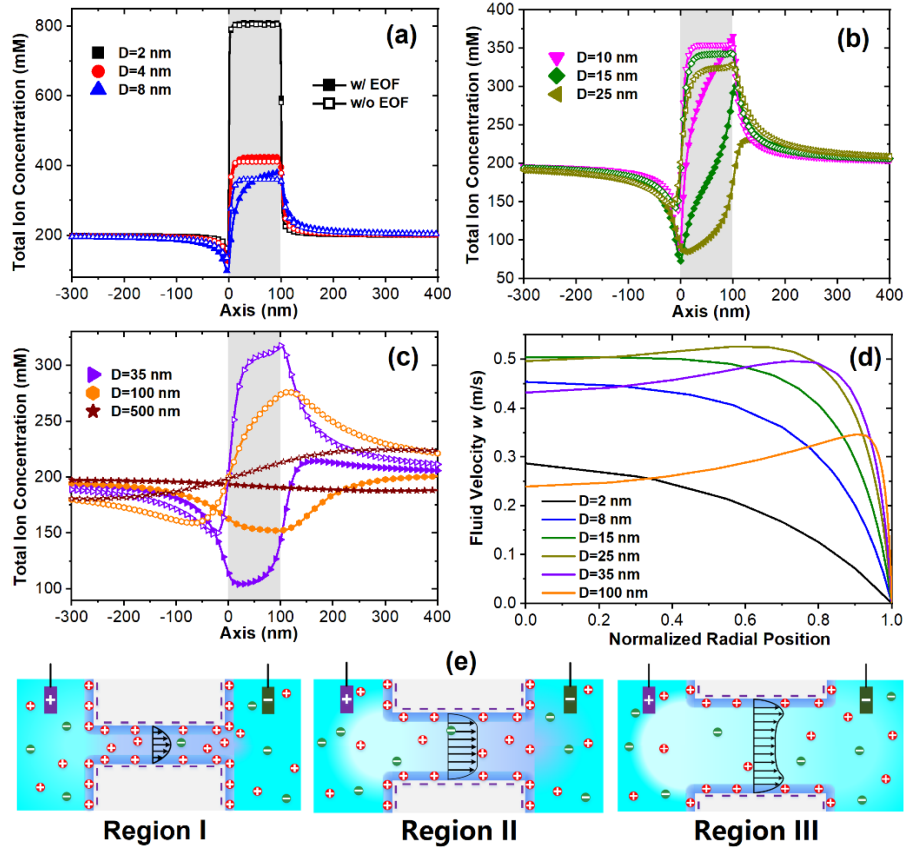

Figure 2 Distributions of the total ion concentration (a-c) along the pore axis and fluid velocity (d) in the radial direction at the center cross-section of nanopores. Pore diameters shown in (a)-(c) are selected from Regions I-III, respectively. (e) Illustrations of EOF in nanopores with various diameters in the three regions defined in Figure 1a.

Besides the convection effect of EOF on ionic transport, the ionic current through nanopores can be influenced by the salt concentration inside nanopores. In Figure 2, the ionic concentration and EOF flow are investigated to provide the detailed mechanism of EOF in the modulation of ionic current under different pore sizes. Following the three regions defined in Figure 1a, three different mechanisms are provided.

For the pore size in Region I, significant overlapping of EDLs induces the serious accumulation of counterions, which results in the high ionic selectivity of nanopores.[18] As plotted in Figure 2a, the total ion concentrations are much higher than the bulk, which have no obvious dependence on the appearance of EOF. One interesting phenomenon is that: with the pore size getting larger, $Cl^-$ ions can approach the pore much easier due to the constant thickness of the electric double layer. To compensate for the concentrated $K^+$ ions, the concentration of $Cl^-$ ions also becomes larger than the bulk values (Figure S3). Due to the highly confined space inside the nanopore, weak EOF forms under the large resistance of fluid flow (Figure 2e).[53] While based on the convection effect of EOF on the counterion movement, EOF has a final promotion to the ionic current. In earlier simulations,[34, 54-56] for narrow short nanopores, fluid flow is usually ignored which may result in a large deviation from the actual situation.

In region II, larger pore sizes induce a lower flow resistance. Stronger EOF (Figure 2d) provides a significant enhancement to the transport of counterions. Accompanying the high ionic selectivity of the nanopore, ion concentration polarization (ICP) happens across the nanopore,[3, 9] i.e. ionic depletion and enrichment happen at the pore entrance and exit, respectively. With the diameter increasing from 10 to 25 nm, the ICP phenomenon becomes weaker but the total ion concentration inside the nanopore gets depleted (Figure 2b). The less obvious ICP is attributed to the increased pore size which decreases the accumulation of permeated counterions. Due to the larger flux through wider nanopores, the relatively slower diffusive replenishment of ions induces the depletion inside the nanopore. Under the combined effect of faster fluid flow and enhanced ionic depletion, the suppression of EOF on $Cl^-$ ion current gradually increased. Although EOF promotes the transport of $K^+$ ions, the lower intra-pore concentration weakens the final flux which plays a dominant role after ICP appears. However, in the cases without consideration of EOF, ICP is negligible, which exhibits that EOF is essential to the appearance of ICP.[3] The enhancement of EOF in cation current decreases from 28% to ~0.9%. Combing both contributions from cations and anions, the total current difference reaches the maximum suppression by EOF at $d$ ~25 nm. In this region, the consideration of EOF is crucial because of its fast speed.

In Region III, with the pore size increasing further, EOF speed starts to decrease, and the intra-pore ionic depletion becomes weaker. The ion concentration in both cases becomes similar due to the weak effect from surface charges and EOF under the extremely large pore sizes (Figure 2c). The total ionic current difference gradually decreases to less than 20% with the diameter increasing to 100 nm. However, under similar distributions of ion concentration inside the nanopore, the current difference of individual species is considerable. For nanopores with a diameter of 500 nm, the difference in $Cl^-$ and $K^+$ ion current between both cases can achieve ~63% and ~53%, respectively. This is due to the promotion and suppression of EOF on ionic transport in wide nanopores. Though the EOF speed is low, the large cross-section area enables a significant deviation of ionic flux, which even induces $t_+$ of 0.8 in the presence of EOF at the diameter of 1000 nm. This phenomenon is counterintuitive because usually large nanopores are thought to have weak ionic selectivity to counterions.[57]

To clarify the unusual ionic selectivity of large nanopores, detailed ionic behaviors have been studied with a nanopore of 100 nm in length and 500 nm in diameter. In the center cross-section of the pore, radial distributions of ionic flux, EOF speed, and concentrations of cations and anions are investigated by dividing the cross-section area of the pore into various ring regions as shown in Figure 3. The concentration distributions of cations and anions follow the Poisson-Boltzmann prediction. Within ~2 nm beyond the charged pore walls, dense cations accumulate in the EDLs which induce a much faster flow near the pore wall. Due to the opposite moving direction to EOF of $Cl^-$ ions and the flow resistance, the speed of EOF decreases in the radial direction until reaching a roughly constant value at ~100 nm away from the pore wall. Because EOF promotes cation transport and suppresses that of anions, the main flux difference in the cross-section happens within 100 nm beyond the surface due to the local stronger fluid flow. Based on the significantly prohibited flow of Cl$^-$ ions and promoted flow of K$^+$ ions, a large cation transfer number is obtained.

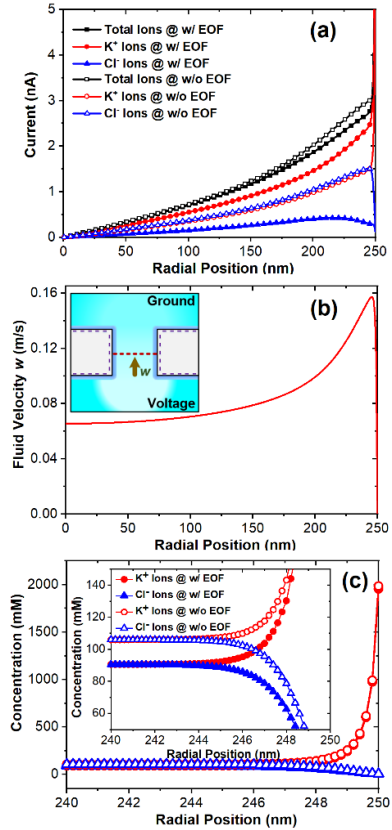

Figure 3 Effect of EOF on ionic behaviors through a nanopore with 100 nm in length and 500 nm in diameter. (a-c) Distributions of ionic flux (a), EOF velocity (b), and concentration of ions (c) in the radial direction. All data were obtained in the center cross-section of the nanopore (inset of b) through the binning method. The fluxes of Cl$^-$ ions are the absolute values.

According to equation 13, the velocity of EOF is positively related to the electric potential and surface charge density, while it is inversely proportional to the pore length, diameter, and solution concentration. Based on our finding of a strong dependence between ionic current and EOF (Figure 1), the influences of EOF on ionic behaviors were investigated under different voltages, surface charge densities, bulk concentrations, and pore dimensions, as shown in Figures 4-7. Figures S4-S14 show corresponding distributions of the total ion concentration along the pore axis and fluid velocity in the radial direction at the center cross-section of nanopores. In addition, we explored the effect of EOF on ionic behaviors under different solution viscosities (Figure S15). Based on the Stokes-Einstein equation, the diffusion coefficient ($D_i$) of ions is adjusted according to the solution viscosity ($\eta$).[58] i.e. $D_i \propto 1/\eta$. The difference in current with or without EOF remains constant with the variation of solution viscosity.

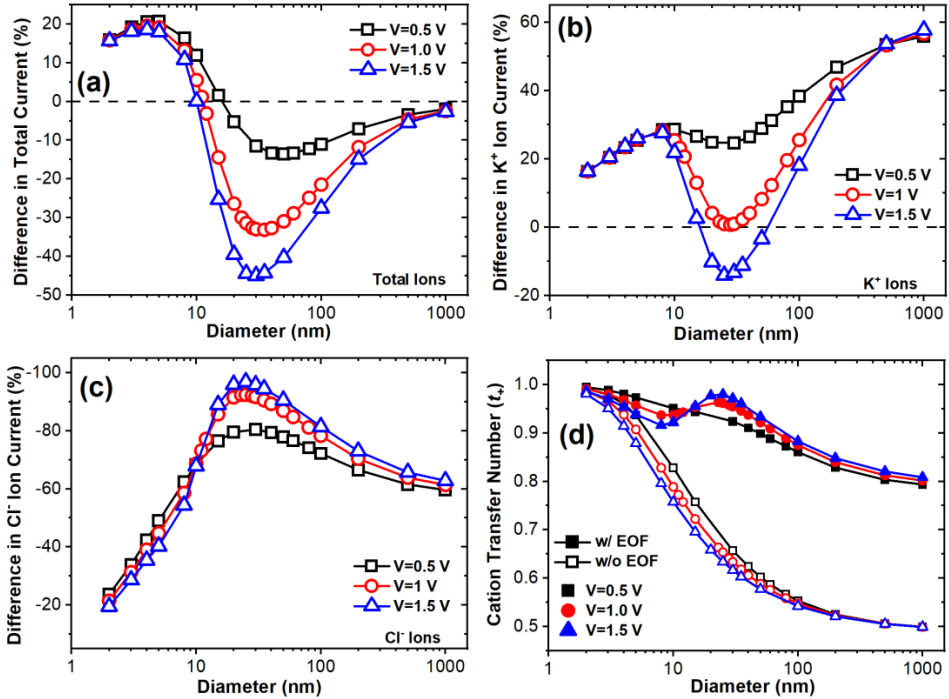

Figure 4 Effect of EOF on ionic behaviors through 100-nm-long nanopores with a variation of the diameter under different voltages. (a)-(c) Difference in ionic current contributed by total ions, K⁺ ions, and Cl⁻ ions. (d) Transfer numbers of cation ($t_+$) under both simulation conditions. The surface charge density was −0.08 C/m². The solution was 100 mM KCl.

Under the constant pore length of 100 nm, the adjustment of applied voltages across the nanopore induces different electric-field strengths, which can modulate the speed of EOF. As shown in Figure 4, the effects of EOF on the total ionic current, as well as the current of cations and anions have similar trends at different voltages. However, in the range of the pore size from 10 to 200 nm, the influence of EOF on the ionic current depends on the applied voltage closely. For the applied voltage of 1.5 V, the difference in total current from both cases with and without EOF can reach ~45% at $d$ ~30 nm. Due to the direct correlation of ICP, ionic depletion, and EOF velocity to the voltage, higher voltage induces stronger EOF and more obvious ionic depletion inside nanopores (Figure S4).[51] Because of the increased intra-pore concentration of ions in the cases without EOF under higher voltages (Figures S4 and S5), the current difference through nanopores with a diameter from 10 to 200 nm exhibit an increasing trend with the applied voltage. Based on the corresponding variation in the current of anions and cations by EOF, the magnitude of the fluctuations in ion selectivity intensifies with the voltage (Figure 4d).

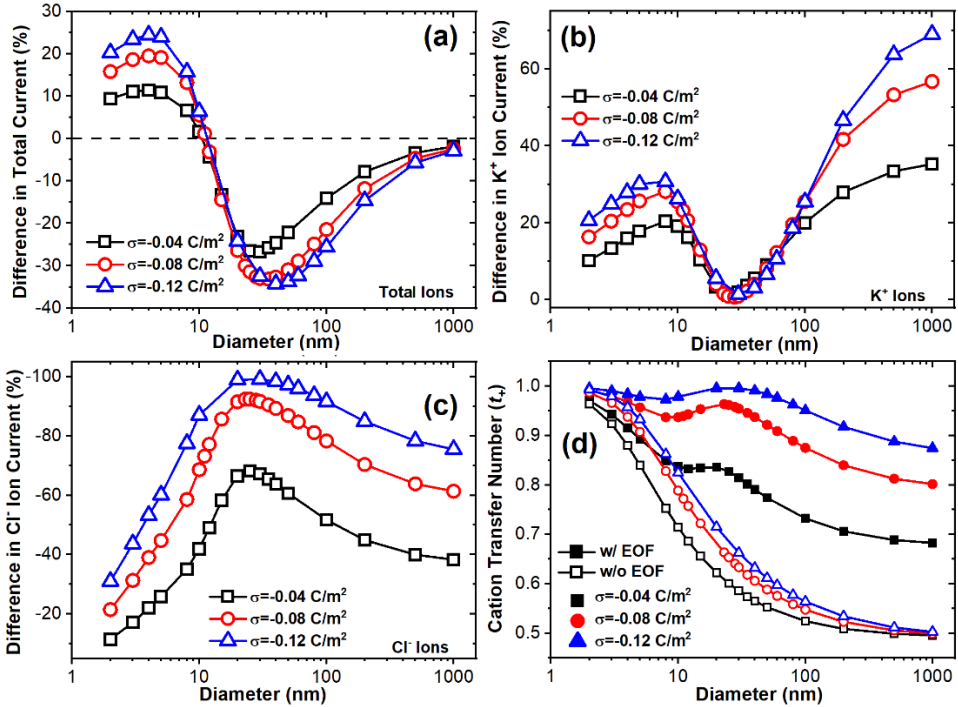

Figure 5 Effect of EOF on ionic behaviors through 100-nm-long nanopores with a variation of the diameter under different surface charge densities. (a)-(c) Difference in ionic current contributed by total ions, $K^+$ ions, and $Cl^-$ ions. (d) Transfer numbers of cation ($t_+$) under both simulation conditions. The solution and voltage were 100 mM KCl, and 1 V, respectively.

The surface charge density is an important parameter of nanopores, which not only determines the ion distribution inside EDLs,[59] but also regulates the velocity of EOF by affecting the wall potential.[1] The effect of EOF on ionic current as a function of the charge density is investigated here as shown in Figure 5. A larger charge density induces a higher concentration of counterions inside the nanopore. Under applied voltages, the stronger transport of cations results in a faster EOF (Figures S6 and S7) and a larger cation transfer number (Figure 5d). In the nanopores with a size located in regions I and III, stronger EOF promotes and inhibits the ionic movement of cations and anions through convection more significantly. While, in region II, the enhancement and suppression of cation current from convection of fluid flow and ICP cancel out, which exhibit no clear dependence on the surface charge density. For a larger surface charge density, EOF should be considered in nanofluidic simulations with nanopores of diameters less than 10 nm even though the space inside nanopores is highly confined.

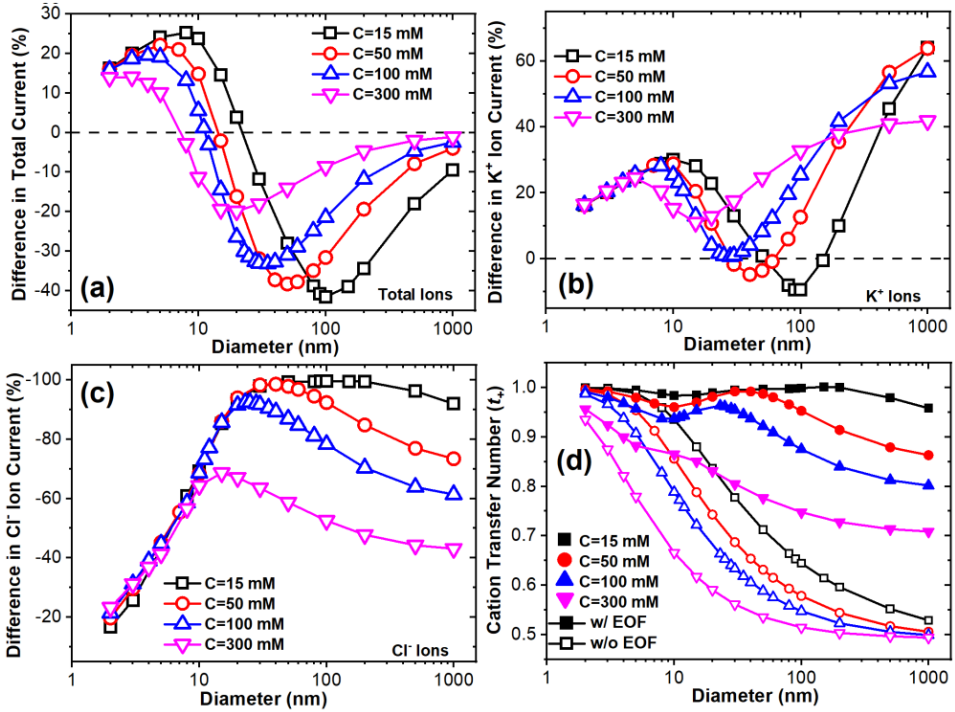

Figure 6 Effect of EOF on ionic behaviors through 100-nm-long nanopores with a variation of the diameter under different salt concentrations. (a)-(c) Difference in ionic current contributed by total ions, K$^+$ ions, and Cl$^-$ ions. (d) Transfer numbers of cation ($t_+$) under both simulation conditions. The surface charge density was −0.08 C/m$^2$. The voltage was 1 V.

Adjustment in solution concentration can cause changes in the bulk conductivity and surface potential,[27] which in turn have a complex effect on the ionic current through nanopores. As shown in Figure 6, the difference in ionic current caused by EOF under different concentrations exhibits similar profiles which exhibited different offsets with the concentration and variation in amplitude. At the pore size of 2 nm, the intra-pore concentration of ions is mainly determined by the surface charge density. In larger nanopores, the ionic conductance inside the nanopore can be divided into two parts, the surface, and bulk conductance. The surface conductance is mainly contributed by the transport of counterions, and the bulk conductance results from the transport of both ions. EOF through the nanopore has convective contribution and suppression by ICP to both surface and bulk conductance. As the pore size increases, the percentage of bulk conductance increases in total conductance. Under a higher concentration, a weaker EOF (Figures S8-S10) and ionic selectivity of the nanopore are induced due to the better electrostatic screening to surface charges. The influence of EOF on both ionic current of cations and anions is weakened in concentrated solutions, which causes the offset to smaller pore sizes and decreased amplitude in the profile of current difference. For simulations with lower salt concentrations, the EOF effect on ionic transport is of great significance.

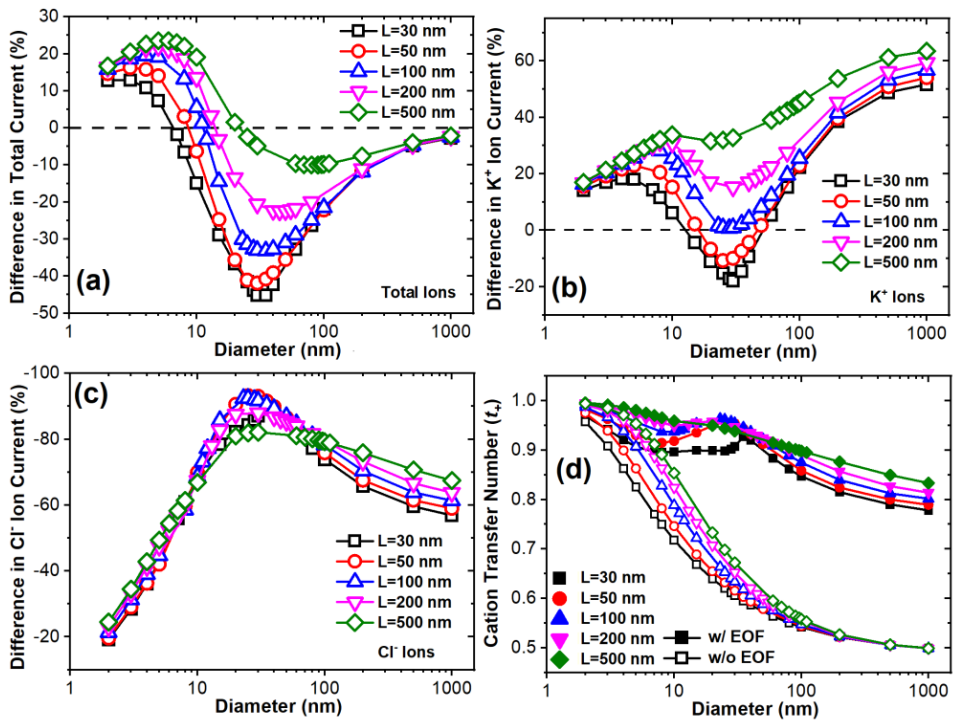

Figure 7 Effect of EOF on ionic behaviors through differently long nanopores with a variation of diameter. (a)-(c) Difference in ionic current contributed by total ions, K$^+$ ions, and Cl$^-$ ions. (d) Transfer numbers of cation ($t_+$) under both simulation conditions. The surface charge density was −0.08 C/m$^2$. The solution and voltage were 100 mM KCl, and 1 V, respectively.

Finally, we explored the influences of EOF on ionic current with different pore lengths. In the nanofluidic system, the whole resistance is composed of pore resistance ($R_p$) and access resistance ($R_{ac}$) which can be described by equation 6 and equation 7, respectively. According to equations 5 to 10, the length of nanopores can affect the pore resistance, the electric field strength, and ion selectivity due to the varied charged area of inner-pore walls.[2] In nanopores with a diameter of less than 10 nm, stronger EOF through shorter nanopores provides obvious convective promotion to the cation transport. In the range of the pore size from ~10 to ~50 nm, because of the more significant ICP under higher electric field strength, the current of K$^+$ ions is suppressed. For nanopores with a constant diameter, the increase in the pore length decreases the electric field strength which induces lower ionic concentrations in the cases without EOF. However, in the simulations with consideration of EOF, the induced slower EOF results in a higher concentration of counterions inside the longer nanopore (Figures S11-S14). With the convective promotion of EOF on the cation transport, the current difference in K$^+$ ion current is enhanced with pore length increasing. For Cl$^-$ ions due to the highly confined space in the nanopores with a diameter from ~10 to 80 nm, gradually slower EOF provides weaker prohibition to their transport with pore length increasing. While, after the pore becomes wider than ~100 nm, access resistance cannot be neglected. Nanopores with the considered lengths share similar EOF velocities. Combined influences of EOF and electrostatic interaction from surface charges inhibit the transport of Cl$^-$ ions more obviously in longer nanopores. For simulations with long nanopores with a small diameter or short nanopores, the fluid flow should be considered in the simulation models.

**IV. CONCLUSIONS**

Through systematical simulations with and without consideration of fluid flow, the influence of EOF on ionic current under various conditions has been investigated through short cylindrical nanopores. Based on the pore size, EOF can affect the ionic current in three different ways. In highly confined nanopores, EOF promotes the ionic current through convective promotion to the transport of counterions i.e. the main current carries. In nanopores of a medium size ranging from ~10 to 30 nm, strong EOF decreases the current significantly due to the induced ionic depletion inside the nanopores. For the nanopores with large diameters, EOF enables the nanopore with considerable ionic selectivity to counterions based on the convective promotion and inhibition from EOF on the transport of counterions and anions. As the pore size increases further, though the influence of EOF on the total current diminishes, its effect on cation/anion current is still considerable. Considering the dependence of EOF speed on the pore dimension and applied conditions, the influences of EOF on the ionic current are systematically explored by varying the applied voltage, surface charge density, salt concentration, and pore length. Simulation results exhibit that in the cases with strong EOF, the fluid flow should be considered in the simulation models. This work provides a comprehensive understanding of the EOF on the ionic current through nanopores and useful guidance for the conduction of continuum simulations.

**SUPPLEMENTARY MATERIAL**

See supplementary material for simulation details, and additional simulation results.

**ACKNOWLEDGMENTS**

This work was supported by the National Science Foundation of China (52105579), the Natural Science Foundation of Shandong Province (ZR2020QE188), the Basic and Applied Basic Research Foundation of Guangdong Province (2019A1515110478), the Natural Science Foundation of Jiangsu Province (BK20200234), the Qilu Talented Young Scholar Program of Shandong University, the Open Foundation of Key Laboratory of High-efficiency and Clean Mechanical Manufacture of Ministry of Education, and the Open Foundation of Key Laboratory of Ocean Energy Utilization and Energy Conservation of Ministry of Education (Grant No. LOEC-202109).

**AUTHOR DECLARATIONS**

**Conflict of Interest**

The author has no conflicts to disclose.

**Author Contributions**

Yinghua Qiu: Conceptualization; Resources; Supervision; Methodology (lead); Investigation (equal); Writing – original draft (equal); Writing – review & editing (lead); Funding acquisition. Long Ma: Methodology (equal); Formal analysis; Software; Validation; Writing – original draft (lead); Writing – review & editing (equal).

**DATA AVAILABILITY**

The data that support the findings of this study are available from the corresponding author upon reasonable request.